\documentclass{PoS-hep}
\usepackage{amsmath}
\usepackage{bm}
\usepackage{epsfig}
\usepackage{graphics}
\usepackage{upgreek}
\usepackage{amsfonts}
\usepackage{amsbsy}
\usepackage{amscd}
\usepackage{bbm}
\usepackage{eufrak}
\usepackage{cite}

\renewenvironment{subequations}{%
\refstepcounter{equation}%
\setcounter{parentequation}{\value{equation}}%
  \setcounter{equation}{0}
  \ignorespaces
}{%
  \setcounter{equation}{\value{parentequation}}%
  \ignorespacesafterend
}
\newcommand{\eeq}{\end{equation}}
\newcommand{\beq}{\begin{equation}}
\newcommand{\ba}{\begin{array}}
\newcommand{\ea}{\end{array}}
\newcommand{\bea}{\begin{eqnarray}}
\newcommand{\eea}{\end{eqnarray}}
\newcommand{\baq}{\begin{eqnarray}}
\newcommand{\eaq}{\end{eqnarray}}

\newcommand{\ecs}{\end{cases}}
\newcommand{\bcs}{\begin{cases}}

\newcommand{\beqs}{\begin{subequations}}
\newcommand{\eeqs}{\end{subequations}}
\newcommand{\eec}{\end{center}}
\newcommand{\bec}{\begin{center}}
\newcommand{\eem}{\end{matrix}}
\newcommand{\bem}{\begin{matrix}}
\newcommand{\Eref}[1]{Eq.~(\ref{#1})}
\newcommand{\Sref}[1]{Sec.~\ref{#1}}
\newcommand{\fref}[1]{Fig.~\ref{#1}}
\newcommand{\Tref}[1]{Table~\ref{#1}}
\newcommand{\cref}[1]{Ref.~\cite{#1}}

\newcommand{\etal}{{\it et al.\/}}

\newcommand\eqs[2]{Eqs.~(\ref{#1}) and (\ref{#2})}

\newcommand\eqss[3]{Eqs.~(\ref{#1}), (\ref{#2}) and (\ref{#3})}

\newcommand{\sFref}[2]{Fig.~\ref{#1}-{\small\sf ({#2})}}

\newcommand{\TeV}{{\mbox{\rm TeV}}}

\newcommand{\GeV}{{\mbox{\rm GeV}}}
\newcommand{\EeV}{{\mbox{\rm EeV}}}
\newcommand{\eV}{{\mbox{\rm eV}}}
\newcommand{\ZeV}{{\mbox{\rm ZeV}}}

\newcommand{\YeV}{{\mbox{\rm YeV}}}

\def\to{\rightarrow}

\def\llgm{\left\lgroup}
\def\rrgm{\right\rgroup}
\def\lf{\left(}
\def\rg{\right)}
\newcommand\vev[1]{\langle {#1} \rangle}
\newcommand{\Gr}{\ensuremath{\widetilde{G}}}
\newcommand{\Yb}{\ensuremath{Y_{B}}}
\newcommand{\Yg}{\ensuremath{Y_{3/2}}}

\newcommand{\Vhi}{\ensuremath{V_{\rm pHI}}}
\newcommand{\Hhi}{\ensuremath{H_{\rm pHI}}}

\newcommand{\mP}{\ensuremath{m_{\rm P}}}
\newcommand{\Mgut}{\ensuremath{M_\mathbb{G}}}

\newcommand{\Gbl}{\ensuremath{\mathbb{G}_{B-L}}}
\newcommand{\Gsm}{\ensuremath{\mathbb{G}_{\rm SM}}}

\newcommand{\lm}{\ensuremath{\lambda_\mu}}

\def\openone{\leavevmode\hbox{\small1\kern-3.8pt\normalsize1}}
\newcommand{\dV}{\ensuremath{\Delta V_{\rm pHI}}}
\newcommand{\Dex}{\ensuremath{\Delta_{\star}}}

\newcommand{\Gsn}{\ensuremath{\widehat{\Gamma}_{\rm \dph}}}
\newcommand{\GNsn}{\ensuremath{\widehat{\Gamma}_{\dph\to N_i^cN_i^c}}}
\newcommand{\Ghsn}{\ensuremath{\widehat{\Gamma}_{\dph\to \hu\hd}}}

\newcommand{\msn}{\ensuremath{\what m_{\rm \dph}}}

\newcommand{\hd}{{\ensuremath{H_d}}}
\newcommand{\hu}{{\ensuremath{H_u}}}

\newcommand\vevi[1]{\langle {#1} \rangle_{\rm I}}

\newcommand{\nmin}{\ensuremath{N_{\rm min}}}
\newcommand{\dphi}{\ensuremath{\what{\delta\phi}}}
\newcommand{\dph}{\ensuremath{\delta\phi}}

\newcommand{\ks}{\ensuremath{k_\star}}
\newcommand{\Ns}{\ensuremath{{N_\star}}}
\newcommand{\ns}{\ensuremath{n_{\rm s}}}

\newcommand{\nst}{\ensuremath{N_{\rm st}}}
\newcommand{\kst}{\ensuremath{K_{\rm st}}}

\newcommand{\as}{\ensuremath{a_{\rm s}}}
\newcommand{\As}{\ensuremath{A_{\rm s}}}

\newcommand{\Ve}{\ensuremath{V}}

\newcommand{\sni}{\ensuremath{N^c_i}}
\newcommand{\ssni}{\ensuremath{\widetilde N_i^c}}

\newcommand{\aS}{\ensuremath{{\rm a}_S}}
\newcommand{\Ald}{\ensuremath{A_\lambda}}
\newcommand{\am}{\ensuremath{{\rm a}_{3/2}}}

\newcommand{\mrh[1]}{\ensuremath{M_{#1N^c}}}
\newcommand{\lrh[1]}{\ensuremath{\lambda_{#1N^c}}}

\newcommand{\mD[1]}{\ensuremath{m_{#1\rm D}}}
\newcommand{\mn[1]}{\ensuremath{m_{#1\nu}}}

\newcommand{\Whi}{\ensuremath{W_{\rm B}}}

\def\ve{\varepsilon}

\def\bbet{{\bar\beta}}
\def\al{{\alpha}}

\def\n{\bar{n}}

\def\th{{\theta}}
\def\thb{{\bar\theta}}
\def\thn{{\theta_{\Phi}}}

\newcommand{\Trh}{\ensuremath{T_{\rm rh}}}
\newcommand{\sg}{\ensuremath{\phi}}

\newcommand{\ld}{\ensuremath{\lambda}}
\newcommand{\ldu}{\ensuremath{\uplambda}}
\newcommand{\Ld}{\ensuremath{\Lambda}}
\newcommand{\kp}{\ensuremath{\kappa}}

\newcommand{\sgx}{\ensuremath{\phi_\star}}
\newcommand{\sgf}{\ensuremath{\phi_{\rm f}}}

\newcommand{\what}{\ensuremath{\widehat}}
\newcommand{\wtilde}{\ensuremath{\widetilde}}

\newcommand{\tks}{\ensuremath{\widetilde K_{\rm I}}}

\newcommand{\se}{\ensuremath{\widehat{\phi}}}
\newcommand{\sex}{\ensuremath{\widehat{\phi}_\star}}
\newcommand{\sef}{\ensuremath{\widehat{\phi}_{\rm f}}}

\newcommand{\mgr}{\ensuremath{m_{3/2}}}

\def\Kap{K\"{a}hler potential}
\def\Km{K\"{a}hler manifold}
\def\Kaa{K\"{a}hler~}

\newcommand{\diag}{\ensuremath{{\sf diag}}}

\newcommand{\fp}{\ensuremath{f_{p}}}
\newcommand{\fps}{\ensuremath{f_{p\star}}}

\newcommand{\nsu}{\ensuremath{{N_X}}}

\newcommand{\fr}{\ensuremath{f_{\rm T}}}
\newcommand{\frs}{\ensuremath{f_{\rm T\star}}}

\newcommand{\phc}{\ensuremath{\Phi}}
\newcommand{\phcb}{\ensuremath{\bar\Phi}}
\newcommand\mtta[4]{\mbox{
$\llgm\bem #1 &#2 \cr #3& #4\eem\rrgm$}}

\newcommand{\bdhh}{{\ensuremath{\normalsize I{\kern-2.9pt H}}}}

\def\act{{\sf\small ACT}}

\def\fhi{{$p$HI}}

\def\actc{{\sf\small P-ACT-LB-BK18}}

\title{\boldmath  ACT-Consistent $B-L$ Higgs Inflation in Supergravity}

\ShortTitle{ACT-Consistent $B-L$ HI in SUGRA}

\author{\speaker{C. Pallis}\\
School of Technology,  \\ Aristotle University of Thessaloniki,\\
Thessaloniki, GR-541 24 GREECE\\
       E-mail: \email{kpallis@auth.gr}}

\abstract{We consider a renormalizable extension of the minimal
supersymmetric standard model (MSSM) endowed by an R and a gauged
$B - L$ symmetry. The model incorporates chaotic inflation driven
by a quartic potential, associated with the Higgs superfields
which lead to a spontaneous breaking of $U(1)_{B-L}$. Consistency
with the ACT data is achieved by considering a fractional
shift-symmetric \Kap\ which includes two free parameters $(p,N)$
constrained in the ranges $1.355\leq p\leq6.7$ and
$6\cdot10^{-5}\leq N\leq0.7$. An explanation of the $\mu$ term of
the MSSM is also provided, under the condition that a related
parameter in the superpotential is somewhat small. Baryogenesis
occurs via non-thermal leptogenesis which is also realized by the
inflaton's decay to the lightest and/or next-to-lightest
right-handed neutrinos for normal ordered light neutrino masses.
\\ \\
{\sl\bfseries Published in}~~{PoS  CORFU {\bf 2025}, 139 (2026)}.
}

\FullConference{Corfu Summer Institute 2025 "School and Workshops
on Elementary Particle Physics and Gravity" (CORFU2025)
24 August - 3 September, 2025\\
Corfu, Greece}

\begin{document}

\section{Introduction}

We focus on a variant of T-model inflation \cite{tmd, sor, tmhi}
called T$_p$-model \emph{inflation} \cite{etmd} which employs as
inflaton the radial component of the Higgs superfields associated
with a $B-L$ phase transition. For this reason we call our model
T$_p$-model \emph{Higgs inflation} ({\sf\small HI}) or for short
{\sf\small pHI} \cite{phi}. We embed it in a complete particle
framework presented in Sec.~\ref{fhim}. The inflationary part of
this context is described in Sec.~\ref{fhi}. Then, in
Sec.~\ref{secmu}, we explain how the $\mu$ term of the
\emph{minimal supersymmetric standard model} ({\sf\small MSSM}) is
obtained and we outline how the observed \emph{baryon asymmetry of
the universe} ({\small\sf BAU}) is generated via \emph{non-thermal
leptogenesis} ({\sf\small nTL}). Our conclusions are summarized in
Sec.~\ref{con}. Throughout the text, the subscript of type $,z$
denotes derivation \emph{with respect to} ({\sf\small w.r.t}) the
field $z$ and charge conjugation is denoted by a star. Unless
otherwise stated, we use units where $\mP = 2.43\cdot
10^{18}~\GeV$ is taken unity.

\section{Model Description}\label{fhim}

We focus on an elementary \emph{Grand Unified Theory} ({\sf \small
GUT}) based on the gauge group $$\Gbl=\Gsm\times
U(1)_{B-L}~~\mbox{with}~~\Gsm= SU(3)_{\rm C}\times SU(2)_{\rm
L}\times U(1)_{Y}$$ being the gauge group of the standard model
and $B$ and $L$ denoting the baryon and lepton number
respectively. The model disposes also three global $U(1)$
symmetries from which the $U(1)_R$ symmetry plays a crucial role
in our construction. The particle content of the model is shown in
\Tref{tab1}. The $i$th generation $SU(2)_{\rm L}$ doublet
left-handed quark and lepton superfields are denoted by $Q_i$ and
$L_i$ respectively, whereas the $SU(2)_{\rm L}$ singlet antiquark
[antilepton] superfields by $u^c_i$ and ${d_i}^c$ [$e^c_i$ and
$\sni$] respectively. The electroweak Higgs superfields which
couple to the up [down] quark superfields are denoted by $\hu$
[$\hd$]. The breaking of $U(1)_{B-L}$ is implemented by the Higgs
superfields $\bar\Phi$ and $\Phi$, which contain the inflaton too,
whereas $S$ is named stabilizer and plays an auxiliary role.

In \Sref{fhim1} and \ref{fhim1} below we specify the building
blocks of our model.

\subsection{Superpotential}\label{fhim1}

The superpotential of our model naturally splits into two parts:
\beq W=W_{\rm MSSM}+\Whi,\>\>\>\mbox{where}\label{Wtot}\eeq
\paragraph{\sf\small (a)} $W_{\rm MSSM}$ is the part of $W$ which contains the
usual terms -- except for the $\mu$ term -- of MSSM, supplemented
by the last term which provides Dirac masses to neutrinos
\beqs \beq W_{\rm MSSM} = h_{ijD} {d}^c_i {Q}_j \hd + h_{ijU}
{u}^c_i {Q}_j \hu+h_{ijE} {e}^c_i {L}_j \hd+ h_{ijN} \sni L_j \hu.
\label{wmssm}\eeq

\paragraph{\sf\small (b)} $\Whi$ is the part of $W$ (``B''eyond that of MSSM) which is relevant for
pHI, the generation of the $\mu$ term of MSSM and the Majorana
masses for neutrinos. It takes the form
\beq\label{Whi} \Whi= \ld S\lf \bar\Phi\Phi-M^2/2\rg+\lm
S\hu\hd+\lrh[i]\phcb N^{c2}_i\,. \eeq\eeqs
The imposed $U(1)_R$ symmetry ensures the linearity of $\Whi$
w.r.t $S$. The first two terms support the implementation of pHI.
The third term plays a key role in the resolution of the $\mu$
problem of MSSM. The fourth term in \Eref{Whi} provides the
Majorana masses for the $\sni$'s and assures the decay of the
inflaton to $\ssni$, whose subsequent decay can activate nTL  --
see \Sref{secmu2}.

\renewcommand{\arraystretch}{1.1}

\begin{table}[!t]
\begin{center}
\begin{tabular}{|c|c|c|c|c|}\hline
{\sc Superfields}&{\sc Representations}&\multicolumn{3}{|c|}{\sc
Global Symmetries}\\\cline{3-5}
&{\sc under $\Gbl$}& {\hspace*{0.3cm} $R$\hspace*{0.3cm} }
&{\hspace*{0.3cm}$B$\hspace*{0.3cm}}&{$L$} \\\hline\hline
\multicolumn{5}{|c|}{\sc Matter Fields}\\\hline
{$e^c_i$} &{$({\bf 1, 1}, 1, 1)$}& $1$&$0$ & $-1$ \\
{$N^c_i$} &{$({\bf 1, 1}, 0, 1)$}& $1$ &$0$ & $-1$
 \\
{$L_i$} & {$({\bf 1, 2}, -1/2, -1)$} &$1$&{$0$}&{$1$}
\\
{$u^c_i$} &{$({\bf 3, 1}, -2/3, -1/3)$}& $1$  &$-1/3$& $0$
\\
{$d^c_i$} &{$({\bf 3, 1}, 1/3, -1/3)$}& $1$ &$-1/3$& $0$
 \\
{$Q_i$} & {$({\bf \bar 3, 2}, 1/6 ,1/3)$} &$1$ &$1/3$&{$0$}
\\ \hline
\multicolumn{5}{|c|}{\sc Higgs Fields}\\\hline
{$\hd$}&$({\bf 1, 2}, -1/2, 0)$& {$0$}&{$0$}&{$0$}\\
{$\hu$} &{$({\bf 1, 2}, 1/2, 0)$}& {$0$} & {$0$}&{$0$}\\
\hline
{$S$} & {$({\bf 1, 1}, 0, 0)$}&$2$ &$0$&$0$  \\
{$\Phi$} &{$({\bf 1, 1}, 0, 2)$}&{$0$} & {$0$}&{$-2$}\\
{$\bar \Phi$}&$({\bf 1, 1}, 0,-2)$&{$0$}&{$0$}&{$2$}\\
\hline\end{tabular}
\end{center}
\caption[]{\sl \small Representations under $\Gbl$ and the extra
global charges of the superfields of our model.}\label{tab1}
\end{table}
\renewcommand{\arraystretch}{1.}

\subsection{\Kaa\ Potential}\label{fhim2}

pHI is feasible if $\Whi$ cooperates \cite{phi} with the following
\Kap
\beq K= \wtilde K_{\rm I}(\phcb,\Phi) +\kst(X^\al),
\label{tkis}\eeq
which includes two contributions: the first one is devoted to
$\phcb-\phc$ system and the second one to $X^\al=S,\hu,\hd,\ssni,
\wtilde L_i$. In particular,
\beqs\beq \kst=\nst\ln\lf1+{|X^\al|^2/\nst}\rg, \label{kst}\eeq
where $0<\nst<6$ and the complex scalar components of the Higgs
superfields are denoted by the same superfield symbol whereas
these of $\sni$ and $L_i$ by $\ssni$ and $\wtilde L_i$
respectively. $\kst$ parameterizes the compact \Kaa manifold
$SU(15)/U(1)$ with constant scalar curvature
$14\cdot15/\nst=210/\nst$ \cite{su11} and successfully stabilizes
$X^\al$ along the inflationary path -- see below -- without
invoking higher order terms. On the other hand, $\wtilde K_{\rm
I}$ includes two terms
\beq \wtilde K_{\rm I}=K_{\rm I}+K_{\rm sh},\label{tki}\eeq
from which $K_{\rm I}$ is real and fractional -- for an
alternative choice see \cref{phi} -- whereas $K_{\rm sh}$ includes
an holomorphic (and an anti-holomorphic) part rendering $\wtilde
K_{\rm I}$ shift-symmetric. Namely,
\beq K_{\rm
I}=\frac{N}{\left(1-|\phc|^2-|\phcb|^{2}\right)^p}~~\mbox{and}~~K_{\rm
sh}=-\frac{N}{2}
\frac{\lf1-2\phcb\phc\rg^p+\lf1-2\phcb^*\phc^{*}\rg^p}
{\left(1-2\phcb\phc\right)^p\left(1-2\phcb^*\phc^{*}\right)^p}.
\label{kksh} \eeq\eeqs
They are defined for $N>0$, $\phcb\phc<1/2$ and
$|\phc|^2+|\phcb|^{2}<1$. $\wtilde K_{\rm I}$ is invariant under
$\Gbl$ and the interchange $\phc\leftrightarrow\phcb$ and
introduces two free parameters $(p,N)$. It parameterizes
hyperbolic \Km\ but without constant curvature, unlike the case of
T-model HI \cite{sor,tmhi}.

\section{Inflationary Scenario}\label{fhi}

The salient features of our inflationary scenario are studied at
tree level in \Sref{fhi1} and at one-loop level in \Sref{fhi11}.
We then present its predictions in \Sref{fhi3}, calculating a
number of observable quantities introduced in Sec.~\ref{fhi2}.

\subsection{Inflationary Potential}\label{fhi1}

We express $\Phi, \bar\Phi$ and $X^\al= S,\hu,\hd,\ssni,\wtilde
L_i$ according to the parametrization
\beq\label{hpar} \Phi=\sg e^{i\th} \cos\thn,~~\bar\Phi={\sg
e^{i\thb}}\sin\thn~~\mbox{and}~~X^\al= ({x^\al +i\bar
x^\al})/{\sqrt{2}},\eeq
where $0\leq\thn\leq\pi/2$. We then find the SUGRA scalar
potential $\Ve$ via the formula
\beq \Ve=\Ve_{\rm F}+ \Ve_{\rm D}\>\>\>\mbox{with}\>\>\> \Ve_{\rm
F}=e^{K}\left(K^{\al\bbet}D_\al \Whi D^*_\bbet \Whi^*-3{\vert
\Whi\vert^2}\right) \>\>\mbox{and}\>\>\Ve_{\rm D}=
{1\over2}g^2{\rm D}_{B-L}^2,\label{Vsugra} \eeq
where $K_{\al\bbet}=K_{,Z^\al Z^{*\bbet}}$,
$K^{\al\bbet}K_{\al\bar\gamma}=\delta^{\bbet}_{\bar \gamma}$,
$D_\al W=W_{,Z^\al} +K_{,Z^\al}W$ and ${\rm D}_{B-L}=Z_\al\lf
B-L\rg K^\al$. Also $g$ is the gauge coupling constant (considered
unified), ${\rm D}_{B-L}$ takes the form
\beq {\rm D}_{B-L}= Np\lf|\phc|-|\phcb|\rg/
2\lf1-|\phc|^2-|\phcb|^2\rg^{p+1}\eeq
(for $\ssni=\wtilde L_i=0$) and suggests that the inflaton can be
identified with the component $\sg$ in \Eref{hpar} which assures
$|\phc|=|\phcb|$. Therefore, \fhi\ can take place along the D-flat
direction
\beq \label{inftr} \vevi{x^\al}=\vevi{\bar
x^\al}=\vevi{\th}=\vevi{\thb}=0\>\>\>\mbox{and}\>\>\>\vevi{\thn}={\pi/4},\eeq
where the symbol $\vevi{Q}$ represents values of $Q$ during \fhi.
The inflationary potential $\Vhi$ can be derived as follows
\beq \label{Vhi} \Vhi=\vevi{\Ve}= \vevi{e^{K}K^{SS^*}} |W_{{\rm
B},S}|^2={\ld^2(\sg^2-M^2)^2}/{4}.\eeq

To determine the canonically normalized fields, we note that
$\vevi{K_{\al\bbet}}$ takes the form
\beq \label{Kab} \vevi{K_{\al\bbet}}=\diag\lf
\vevi{M_{\phcb\phc}},\underbrace{1,...,1}_{14~\mbox{\small
elements}}\rg~~\mbox{with}~~
\vevi{M_{\phcb\phc}}=\frac{pN}{\fr^{2+p}}\mtta{\kappa}{\bar\kappa}{\bar\kappa}{\kappa}
~~\mbox{and}~~K_{X^\al X^{*\bar\al}}=1.\eeq
Here $\kp=\fr+fp$, with $\fr=1-\sg^2$ and $\fp=1+p\sg^2$, and
$\bar\kp=(1+p)\sg^2$. Upon diagonalization of
$\vevi{M_{\phcb\phc}}$ we find its eigenvalues which are
\beq
\label{kpm}\kp_+=\kp=pN\fp/\fr^{p+2}~~\mbox{and}~~\kp_-=pN/\fr^{p+1}.\eeq
Inserting \eqs{hpar}{Kab} into the kinetic term of the SUGRA
action, $K_{\al\bbet}\dot Z^\al\dot Z^\bbet$, we can specify the
canonically normalized fields, denoted by hat, as follows
\beq \label{VJe} \frac{d\se}{d\sg}=J,~~\widehat{\theta}_\pm
=\sqrt{\kp_\pm}\sg\theta_\pm,~~ \widehat{\theta}_\Phi =
\sqrt{2\kp_-}\sg\lf\theta_\Phi-\frac{\pi}{4}\rg~~\mbox{and}~~(\what{x}^{\al},\what{\bar
x}^{\al})=(x^\al,\bar x^\al)\,,\eeq 
where $J=\sqrt{2\kp_+}$ and
$\th_{\pm}=\lf\bar\th\pm\th\rg/\sqrt{2}$. As we show below, the
masses of the scalars besides $\se$ during pHI are heavy enough
such that the dependence of the hatted fields on $\sg$ does not
influence their dynamics.

\begin{figure}[!t]\vspace*{-.18in}
\hspace*{-.12in}
\begin{minipage}{8in}
\epsfig{file=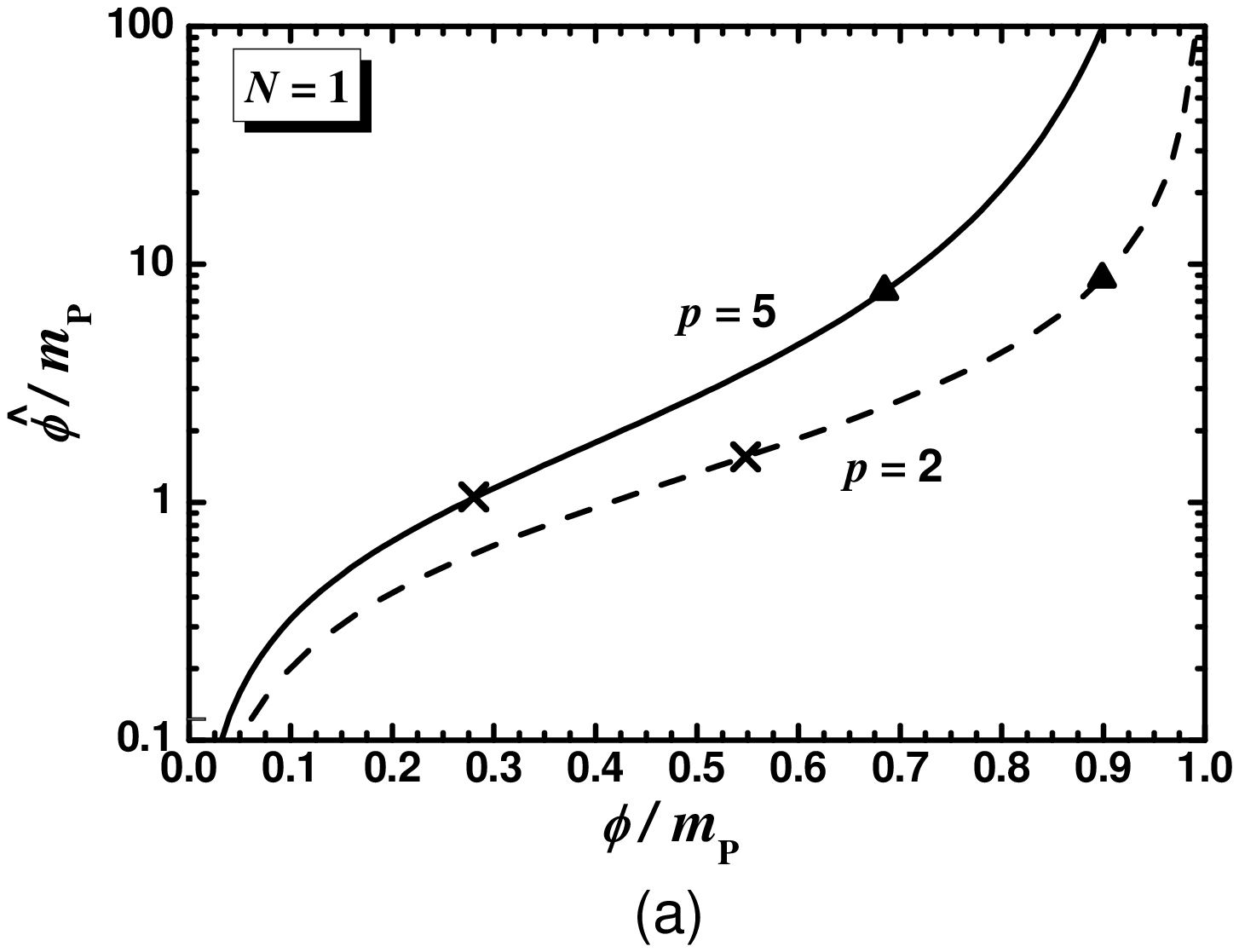,height=3.6in,angle=-90} \hspace*{-1.3cm}
\epsfig{file=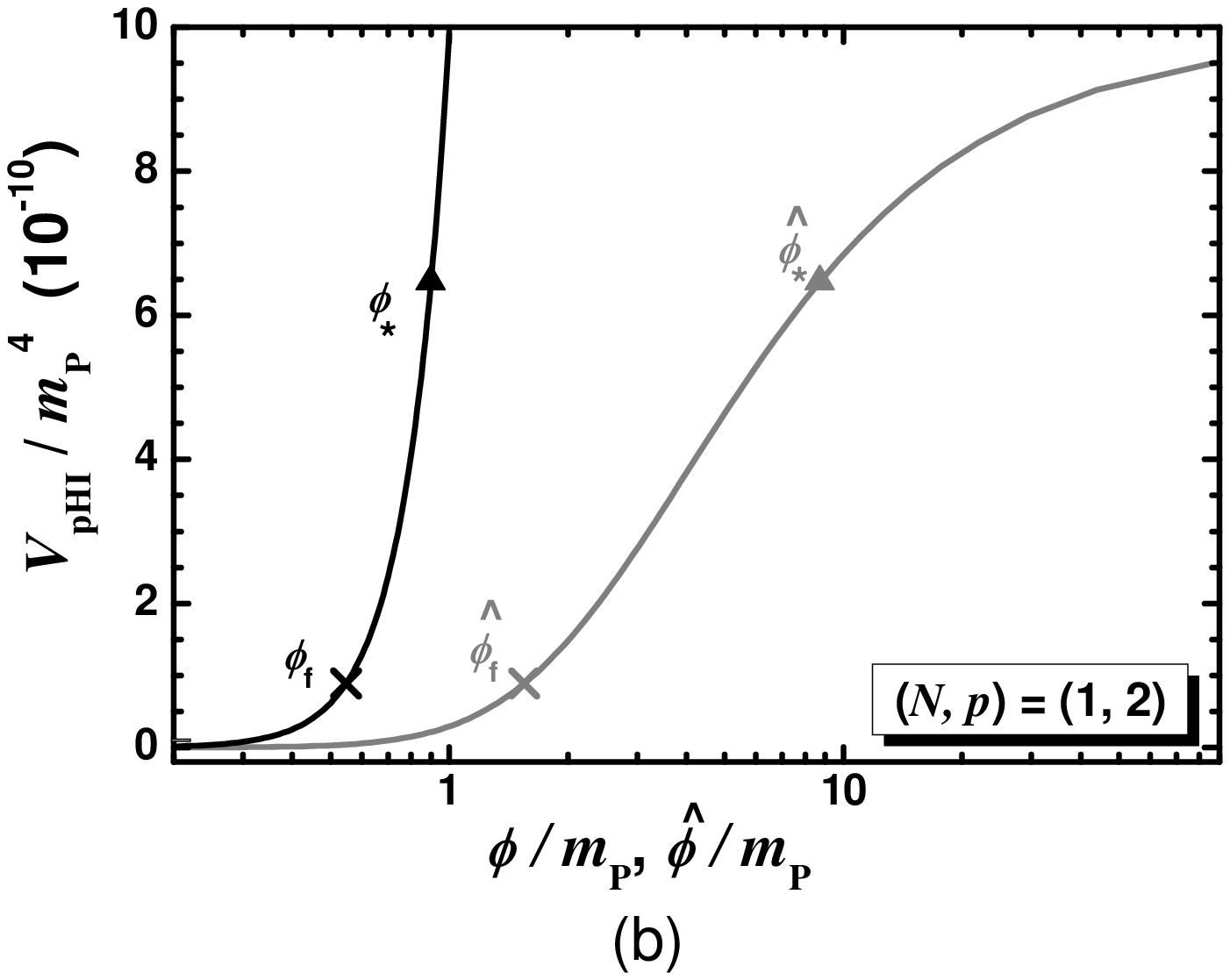,height=3.6in,angle=-90} \hfill
\end{minipage}
\hfill \caption{\sl\small {\sffamily\small (a)} Canonically
normalized inflaton $\se$ as a function of $\sg$ for $N=1$ and
$p=5$ (solid line) or $p=2$ (dashed line); {\sffamily\small (b)}
inflationary potential $\Vhi$ for $(p,N)=(1,2)$ as a function of
$\sg$ (black line) and $\se$ (gray line). Values corresponding to
$\sgx$, $\sgf$, $\sex$ and $\sef$ are also depicted in both
panels.}\label{fig1}
\end{figure}

Taking into account the results in \eqs{Vhi}{VJe} we depict $\se$
as a function of $\sg$ for $N=1$ and $p=5$ (solid line) or $p=2$
(dashed line) in \sFref{fig1}{a} and $\Vhi$ for $(p,N)=(2,1)$ as a
function of $\sg$ (black line) and $\se$ (gray line) in
\sFref{fig1}{a}. From the former plot, we can remark that $\se$
increases w.r.t $\sg$ rendering \fhi\ possible even for $\sg<1$,
whereas from the latter we see that $\Vhi$ as a function of $\sg$
has the well-known parabolic-like slope but as a function of $\se$
experiences a stretching for $\se>1$ rendering it suitable for pHI
-- cf. \cref{tmd,etmd,sor}.

\subsection{Stability and one-Loop Radiative Corrections}\label{fhi11}

We here demonstrate that the inflationary direction in
\Eref{inftr} is stable w.r.t the fluctuations of the non-inflaton
fields. To this end, we construct the mass-squared spectrum of the
scalars taking into account the canonical normalization of the
various fields in \Eref{VJe}. We find the expressions of the
masses squared $\what m^2_{z^\al}$ (with
$z^\al=\theta_+,\theta_\Phi, x^\al$ and $\bar x^\al$) arranged in
\Tref{tab3}. The various unspecified there eigenstates are defined
as follows
\beqs\beq  h_\pm=(h_u\pm{h_d})/\sqrt{2},~~~ \widehat{\bar
h}_\pm=(\widehat{\bar h}_u\pm\widehat{\bar
h}_d)/\sqrt{2}~~~\mbox{and}~~~\what \psi_\pm
=(\what{\psi}_{\Phi+}\pm \what{\psi}_{S})/\sqrt{2}, \eeq
where the (unhatted) spinors $\psi_\Phi$ and $\psi_{\bar\Phi}$
associated with the superfields $\Phi$ and $\bar\Phi$ are related
to the normalized (hatted) ones in \Tref{tab3} as follows
\beq \label{psis}
\what\psi_{\Phi\pm}=\sqrt{\kp_\pm}\psi_{\Phi\pm}~~~
\mbox{with}~~~\psi_{\Phi\pm}=(\psi_\Phi\pm\psi_{\bar\Phi})/\sqrt{2}\,.
\eeq\eeqs

From \Tref{tab3} it is evident that $0<\nsu\leq6$ assists us to
achieve $m^2_{{s}}>\Hhi^2=\Vhi/3$ -- in accordance with the
results of \cref{su11} -- and also enhances the ratios
$m^2_{X^{\gamma}}/\Hhi^2$ for $X^{\gamma}=\hu,\hd,\ssni,\wtilde
L_i$ w.r.t the values that we would have obtained, if we had used
just canonical terms in $K_{\rm st}$ -- see \Eref{kst}. On the
other hand, $m^2_{h-}>0$ requires
\beq \label{lmb} \lm<\ld(1+\nst)\sgf^2/2\nst, \eeq
where $\sgf$ is the $\sg$ value at the end of \fhi\ -- see
\Sref{fhi2} below. The upper bound above numerically equals to
$2\cdot10^{-5}$ for $\nst=1$. Despite its low value, $\lm$ can
become compatible with $\mu/\mgr\sim1$ as we show in \Sref{secmu1}
-- here $\mgr$ is the $\Gr$ mass. In \Tref{tab3} we display also
the mass $M_{BL}$ of the gauge boson which becomes massive having
`eaten' the Goldstone boson $\th_-$. This signals the fact that
$\Gbl$ is broken during pHI and so no cosmological defects (cosmic
strings in our case) are produced.

The derived mass spectrum can be employed in order to find the
one-loop radiative corrections $\dV$ to $\Vhi$. Considering SUGRA
as an effective theory with cutoff scale equal to $\mP$, the
well-known Coleman-Weinberg formula can be employed
self-consistently taking into account only the masses which lie
well below the UV cut-off scale $\mP$, i.e., all the masses
arranged in \Tref{tab3} besides $M_{BL}$ and $m_{\th_\Phi}$ -- cf.
\cref{phi, unv, jhep}. The resulting $\dV$ takes the form
\beq\label{dV} \dV=\frac{1}{64\pi^2}\sum_\al {{\rm N}_\al}
m_\al^4\ln \frac{m_\al^2}{\Ld_{\rm CW}^2}-6M_{iN^c}^4\ln
\frac{M_{iN^c}^2}{\Ld_{\rm CW}^2}~~\mbox{where}~~\bcs
\al&=\{\theta+,s,h\pm,i\tilde \nu^c,i\tilde
l,\psi_{\Phi\pm}\}\\{\rm N}_\al&=\{1,2,8,6,12,-4\}\ecs\eeq
and lets intact our inflationary outputs, provided that the
renormalization-group mass scale $\Lambda_{\rm CW}$, is determined
by requiring $\dV(\sgx)=0$ or $\dV(\sgf)=0$ -- see below.

\renewcommand{\arraystretch}{1.2}
\begin{table}[!t]
\bec\begin{tabular}{|c||c|c|c|c|}\hline {\sc Fields}&{\sc
Eigenstates}& \multicolumn{3}{|c|}{\sc Masses Squared}\\
\hline\hline
%
30 real&$\widehat\theta_{+}$&$m_{\theta+}^2$&
\multicolumn{2}{|c|}{$6\Hhi^2$}\\
scalars&$\widehat \theta_\Phi$ &$m_{
\theta_\Phi}^2$&\multicolumn{2}{|c|}{$M^2_{BL}+6\Hhi^2$} \\
&$s, {\bar{s}}$ &$ m_{
s}^2$&\multicolumn{2}{|c|}{$6\Hhi^2\lf1/\nst+2\fr^{p+2}/pN\sg^2\fp\rg$} \\
& $h_{\pm},{\bar h}_{\pm}$ &
$m_{h\pm}^2$&\multicolumn{2}{c|}{$3\Hhi^2\lf1+1/\nst
\pm{2\lm}/{\ld\sg^2}\rg$}\\
& $\tilde \nu^c_{i}, \bar{\tilde\nu}^c_{i}$ & $ m_{i\tilde
\nu^c}^2$&\multicolumn{2}{c|}{$3\Hhi^2\lf1+1/\nst+8\ld^2_{iN^c
}/\ld^2\sg^2\rg$}\\ & $\tilde l_{i}, \bar{\tilde l}_{i}$ & $
m_{i\tilde l}^2$&\multicolumn{2}{|c|}{$3\Hhi^2(1+1/\nst)$}\\
\hline
1 gauge boson &{$A_{BL}$}&{$M_{BL}^2$}&
\multicolumn{2}{|c|}{$2g^2Np\sg^2/\fr^{p+1}$}\\\hline
$7$ Weyl  & $\what \psi_\pm$ & $m^2_{ \psi\pm}$&
\multicolumn{2}{|c|}{$6\Hhi^2\fr^{p+2}/pN\sg^2\fp$}\\
spinors &{$N_i^c$}& {$
M_{{iN^c}}^2$}&\multicolumn{2}{c|}{$24\Hhi^2\ld^2_{iN^c}/\ld^2\sg^2$}\\
&$\ldu_{BL},
\widehat\psi_{\Phi-}$&$M_{BL}^2$&\multicolumn{2}{|c|}{$2g^2Np\sg^2/\fr^{p+1}$}\\
\hline
\end{tabular}\eec
\hfill \caption{\sl\small The mass squared spectrum of our models
along the path in Eq.~(3.4). }\label{tab3}
\end{table}
\renewcommand{\arraystretch}{1.}

\subsection{Inflationary Observables}\label{fhi2}

A period of slow-roll pHI is controlled by the strength of the
slow-roll parameters
\beq\label{sr}\epsilon = \left(\frac{\Ve_{{\rm
pHI},\se}}{\sqrt{2}\Vhi}\right)^2\simeq\frac{4\fr^{p+2}}{pN
\fp\sg^2}~~\mbox{and}~~\eta = \frac{\Ve_{{\rm pHI},\se\se}}{\Vhi}
\simeq2 \fr^{p+1}\frac{3 + (p-5)\sg^2 -
p(p+4)\sg^4)}{pN\sg^2\fp^2} \cdot\eeq
Expanding $\epsilon$ and $\eta$ for $\sg\ll 1$ we can find that
pHI terminates for $\sg=\sgf$ such that
\beq{\small\sf max}\{\epsilon(\sgf),|\eta(\sgf)|\}=1
~~\Rightarrow~~\sgf\simeq\sqrt{6/(16 + 3Np)}, \label{sgap}\eeq
which increases as $N$ decreases and so pHI can be established
easier for relatively large $N$'s.

The number of e-foldings, $\Ns$, that the pivot scale
$\ks=0.05/{\rm Mpc}$ suffers during pHI can be calculated through
the relation
\begin{equation}
\label{Nhi}  \Ns=\:\int_{\se_{\rm f}}^{\se_\star}\, d\se\:
\frac{\Ve_{\rm pHI}}{\Ve_{\rm pHI,\se}}\simeq \frac{pN\sgx^2}{4
\frs^{1+p}}\>\Rightarrow\>\sgx\simeq \sqrt{1 - \lf\frac{pN}{4
\Ns}\rg^{1/(p+1)}},
\end{equation}
where $\sex$ [$\sgx$] is the value of $\se$ [$\sg$] when $\ks$
crosses the inflationary horizon and we take into account that
$\sgf\ll\sgx$. Apparently, $\sgx<1$ and so our proposal can be
stabilized against corrections from higher order terms of the form
$(\phcb\phc)^\ell$ with $\ell>1$ in $\Whi$ -- see \Eref{Whi}.

The amplitude $\As$ of the power spectrum of the curvature
perturbations generated by $\sg$ at the pivot scale $\ks$ is
estimated as follows
\beq \label{Proba}\sqrt{\As}=\: \frac{1}{2\sqrt{3}\, \pi} \;
\frac{\Ve_{\rm pHI}(\sex)^{3/2}}{|\Ve_{\rm
pHI,\se}(\sex)|}~~\Rightarrow~~
\ld\simeq2^{(n/4+5/2)}\sqrt{3\As}\pi\frs^{p/2+1}/{\sgx^3\sqrt{pN\fps}}.\eeq
The resulting relation yields numerically $\ld\sim10^{-6}$.

At the pivot scale we can also calculate the scalar spectral index
$\ns$, its running $\as$, and the tensor-to-scalar ratio $r$ via
the relations
\beqs\baq \label{ns} && \ns=\: 1-6\epsilon_\star\ +\
2\eta_\star\simeq
1-\frac{p+2}{(p+1)\Ns}\,,~~r=16\epsilon_\star\simeq
\frac{4(4^ppN)^{1/(p+1)}}{(p+1)N_\star^{(p+2)/(p+1)}},\\
&& \label{as} \as =\:{2\over3}\left(4\eta_\star^2-(n_{\rm
s}-1)^2\right)-2\xi_\star\simeq-\frac{p+2}{(p+1)N^2_\star}
~~\mbox{with}~~ \xi={\Ve_{\rm pHI,\widehat\sg} \Ve_{\rm
pHI,\widehat\sg\widehat\sg\widehat\sg}/\Ve_{\rm
pHI}^2}.~~\eaq\eeqs
Here the variables with subscript $\star$ are evaluated at
$\sg=\sgx$. A clear dependence of the observables on $p$ arises
which renders our results friendly to ACT as it is evident for
$p=1$ or $2$  and $\Ns=55$. Note also that the expressions above
for $\ns$ and $\as$ are independent from $N$. This fact, however,
is not totally kept as we show numerically below.


\subsection{Comparison with Observations}\label{fhi3}

The approximate analytic expressions above can be verified by the
numerical analysis of our model. Namely, we apply the accurate
expressions in \eqs{Nhi}{Proba} and confront the corresponding
observables with the requirements \cite{actin}
\begin{equation}
\label{Ntot}
\Ns\simeq61.5+\ln\lf\Vhi(\sgx)^{1/2}/\Vhi(\sgf)^{1/4}\rg~~~\mbox{and}~~~\As^{1/2}\simeq4.618\cdot10^{-5}\,,\eeq
where we consider in the left-hand side relation  an
equation-of-state parameter $w_{\rm rh}\simeq1/3$ correspoding to
quartic potential which approximates \cite{phi} rather well $\Vhi$
for $\sg\ll1$. The constraints above allows us to extract $\ld$
and $\sgx$ for any given $p$, $N$ and $M$. However, if $M\ll\mP$
then $M$ is irrelevant for the inflationary observables. Motivated
by the unification of the gauge coupling constants within MSSM, we
determine $M$ doing the identification
\beq \label{Mg}
\vev{M_{BL}}=\Mgut~~\Rightarrow~~M\simeq{\Mgut}/{g\sqrt{2Np}},\eeq
where we take into account that $\vev{\fr}\simeq1$ and
$\vev{\sg}=M$ in accordance with \Eref{vevs} -- see below. We use
the numerical values $g\simeq0.7$ for the value of the GUT gauge
coupling constant and $\Mgut\simeq2/2.433\cdot10^{-2}$ for the
unification scale. The success of our inflationary setting is
assured if we impose a lower bound on $N$ derived as follows
\beq \label{Nmin} M\lesssim\mP~~\Rightarrow~N\gtrsim
\nmin={\Mgut^2}/{2g^2p\mP^2}.\eeq

Given \Eref{Mg}, our final set of free input parameters is $(p,N)$
and it can be constrained by comparing our results on $\ns$ and
$r$, computed via the definitive relations in \Eref{ns}, against
the combination of the \act\ DR6 with the measurements from other
experiments \cite{actin}. The so-obtained \actc\ data suggests
\begin{equation}  \label{nsdata}
\ns=0.9743\pm0.0068\>\>\mbox{and}\>\>r\leq0.038,
\end{equation}
at 95$\%$ \emph{confidence level} ({\sf\small c.l.}) with
$|\as|\ll0.01$ -- ramifications of the restrictions above for
several inflationary models are recently reviewed in
\cref{actreview}. More precisely, in \sFref{fig2}{a}, we place
over the top of the allowed contours of the \actc\ data in the
$\ns-r$ plane the lines predicted by our model after imposing
\eqss{Ntot}{Mg}{Nmin}. We draw solid, dashed, dot-dashed and
dotted lines for $p=1, 2, 5$ and $10$ respectively and show the
variation of $N$ along each line. For the two latter $p$'s the
corresponding \nmin\ values from \Eref{Nmin} are also displayed.
We clearly see that $\ns$ increases with $p$ and $r$ with $N$ and
so the whole shaded region favored by \actc\ is covered varying
$p$ and $N$. As a bonus, the resulting $r$'s for $(p,N)$'s of
order unity are also testable by the forthcoming experiments
\cite{det}, like {\sc Bicep3}, PRISM and LiteBIRD, searching for
primordial gravity waves since $r\gtrsim0.007$.

\begin{figure}[!t]\vspace*{-.18in}
\hspace*{-.19in}
\begin{minipage}{8in}
\epsfig{file=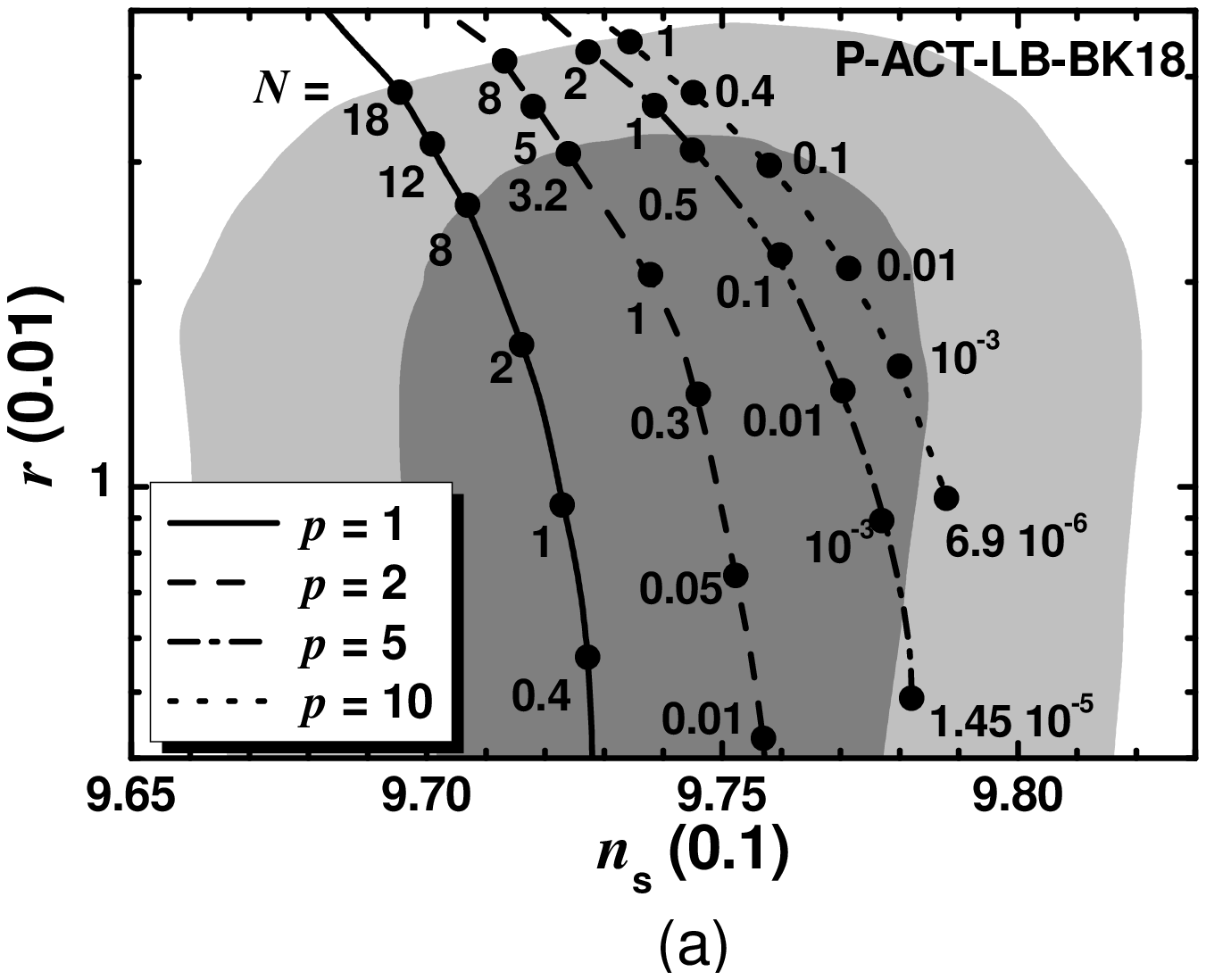,height=3.6in,angle=-90}
\hspace*{-1.3cm}
\epsfig{file=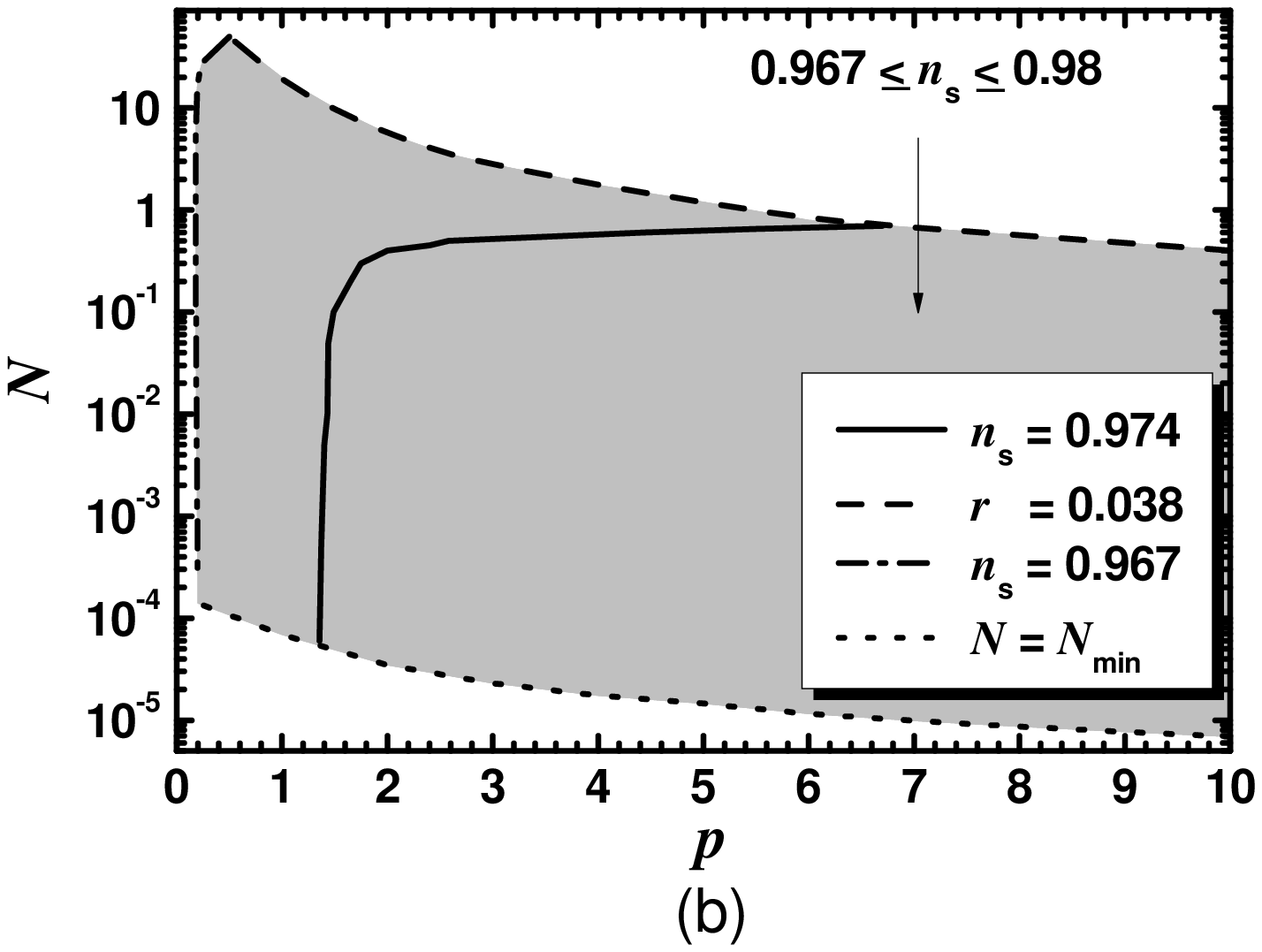,height=3.6in,angle=-90} \hfill
\end{minipage}
\hfill \caption{\sl\small {\sffamily\small (a)} Allowed curves in
the $\ns-r$ plane with the $N$ values indicated along them  -- the
marginalized joint $68\%$ [$95\%$] region from \actc\ data is
depicted by the dark [light] shaded contour; {\sffamily\small (b)}
Allowed (shaded) region as determined by Eq.~(3.17) -- (3.20) in
the $p-N$ plane. The conventions adopted for the various lines are
also shown in both panels.}\label{fig2}
\end{figure}\renewcommand{\arraystretch}{1.}


The variation of the observables in \sFref{fig2}{a} reveals that
our free parameters can be constrained in the $p-N$ plane. The
relevant allowed region is shown in \sFref{fig2}{b} with boundary
curves originated from  {\small\sf (i)} the bound on $r$ in
\Eref{nsdata} (dashed line), {\small\sf  (ii)} the lower bound on
$\ns$ in \Eref{nsdata} (dot-dashed line) and {\small\sf  (iii)}
the lower bound on $N$ in \Eref{Nmin} (dotted line) --  we take by
hand a maximal $p=10$ since no upper bound on $p$ can be inferred
from the corresponding bound on $\ns$ in \Eref{nsdata}. Fixing
$\ns$ to its central value in \Eref{nsdata}, we obtain the solid
line along which we obtain
\beq \label{res1} 1.355\lesssim
p\lesssim6.7~~\mbox{and}~~6\cdot10^{-5}\lesssim N\lesssim0.7\eeq
with $\Ns\simeq(54.7-56.8)$, $\as\simeq -4.7\cdot10^{-4}$ and
$r\gtrsim 2.8\cdot10^{-4}$. The obtained $|\as|$'s remain
negligibly small being, thereby, consistent with its $95\%$ c.l.
allowed margin in \cref{actin}. The closer to unity $\sgx$ is
chosen, the largest $\Ns$ is obtained. To quantify the relevant
tuning we estimate $\Dex=1-\sgx$ which ranges in the interval
$(0.8-28)\%$ for the values in \Eref{res1}. The maximal $\Dex$
values are obtained for the largest $p$'s and $N$'s indicating,
thereby that naturalness prefers $N$ and $p$ values of order
unity.

Note, finally, that the one loop radiative corrections in
\Eref{dV} let immune our results provided we impose one of the two
conditions mentioned below the aforementioned equation. E.g.,
imposing the first of them for $\nst=1$ and $p=2$ we find
$\Ld_{\rm CW}\simeq(4-5)~\YeV$ -- hereafter we restore units,
i.e., we take $\mP=2.43\cdot10^{18}~\GeV$. Under these
circumstances, our inflationary predictions can be exclusively
reproduced by using $\Vhi$ in \Eref{Vhi}.

\section{Post-Inflationary Scenario}\label{secmu}

Two byproducts of our setting is the derivation of the $\mu$ term
of MSSM, as shown in \Sref{secmu1}, and the implementation of
baryogenesis via nTL as described in \Sref{secmu2}.

\subsection{Generation of the $\mu$ Term of MSSM} \label{secmu1}

The origin of the $\mu$ term of MSSM can be explained if we
combine the three first terms of \Eref{Whi} working at the level
of SUSY. In fact, the total low energy potential is
\beq V_{\rm tot}=V_{\rm SUSY}+V_{\rm soft},\label{vtot}\eeq
where the first term includes the SUSY limit $V_{\rm SUSY}$ of
$\Vhi$ in \Eref{Vhi} which is given by
\beqs \beq \label{Vsusy} V_{\rm SUSY}= K_{\rm SUSY}^{\al\bbet}
W_{\rm B\al} W^*_{\rm B\bbet}.\eeq
Here $K_{\rm SUSY}$ is the limit of the $K$ \Eref{tkis} for
$\mP\to\infty$, which is
\beq \label{Kquad}K_{\rm
SUSY}=Np\lf|\phc|^2+|\phcb|^2-\phcb\phc-\phcb^*\phc^*\rg+|X^\al|^2.\eeq
Upon substitution of $K_{\rm SUSY}$ into \Eref{Vsusy} we obtain
\bea \nonumber && V_{\rm SUSY}=\ld^2|\phcb\phc
-M^2/2|^2+\ld^2|S|^2\lf|\phcb|^2+|\phc|^2\rg/Np+\cdots,
\label{VF}\eea\eeqs
where the ellipsis includes terms which vanish at the SUSY vacuum
lying along the D-flat direction $|\phcb|=|\phc|$ with
\beq \vev{X^\al}=0 \>\>\mbox{and}\>\>
|\vev{\Phi}|=|\vev{\bar\Phi}|=M/\sqrt{2}\,.\label{vevs} \eeq
As a consequence, $\vev{\Phi}$ and $\vev{\bar\Phi}$ break
spontaneously $U(1)_{B-L}$ down to $\mathbb{Z}^{B-L}_2$. Since
$U(1)_{B-L}$ is already broken during pHI, no cosmic strings are
formed.

The contributions from the soft SUSY-breaking terms, although
negligible during pHI, since these are much smaller than $\sg$,
may shift slightly $\vev{S}$ from zero in \Eref{vevs}. Indeed, the
relevant potential terms are
\beq V_{\rm soft}= \lf\ld A_\ld S \phcb\phc+\lm A_\mu S \hu\hd +
\ld_{iN^c} A_{iN^c}\phc \widetilde N^{c2}_i- {\rm a}_{S}S\ld M^2/2
+ {\rm h. c.}\rg+ m_{\al}^2\left|X^\al\right|^2, \label{Vsoft}
\eeq
where $m_{\al}, A_\ld, A_\mu, A_{iN^c}$ and $\aS$ are soft SUSY
breaking mass parameters.  Rotating $S$ in the real axis by an
appropriate $R$-transformation, choosing conveniently the phases
of $\Ald$ and $\aS$ so as the potential in \Eref{vtot} to be
minimized -- see \Eref{VF} -- and substituting in $V_{\rm soft}$
$\vev{\phc}$  and $\vev{\phcb}$ from \Eref{vevs} we get
\beqs\beq \vev{V_{\rm tot}(S)}= \ld^2{M^2S^2}/{Np} -2\ld\am\mgr
M^2S,~~\mbox{where}~~\am={(|A_\ld| + |{\rm
a}_{S}|)}/{2\mgr}>0\label{Vol} \eeq
is a parameter of order unity which parameterizes our ignorance
for the dependence of $|A_\ld|$ and $|{\rm a}_{S}|$ on $\mgr$ and
we take into account that $m_S\ll M$. The minimization condition
for the total potential in \Eref{Vol} w.r.t $S$ leads to a non
vanishing $\vev{S}$ as follows
\beq \label{vevS}\frac{d}{d S} \vev{V_{\rm
tot}(S)}=0~~\Rightarrow~~\vev{S}\simeq 2Np\
\am\mgr/{\ld}.\eeq\eeqs
The generated $\mu$ term from the third term in \Eref{Whi} is
\beq\mu =\lm \vev{S} \simeq 2\lm Np\ \am\mgr/{\ld}.\label{mu}\eeq
It can be comparable to $\mgr$ for $p$ and $N$ of order unity, if
$\lm$ is of the same order of magnitude with $\ld$. Since $\ld\sim
10^{-6}$, as inferred by \Eref{Proba}, the required $\lm$ can
become compatible with \Eref{lmb} assuring, thereby, the stability
of the path in \Eref{inftr} during \fhi.

\subsection{Baryogenesis, $\Gr$ Abundance and Neutrino
Masses}\label{secmu2}

When pHI is over, the inflaton continues to roll down towards the
SUSY vacuum, \Eref{vevs}. Soon after, it settles into a phase of
damped oscillations around the minimum of $\Vhi$. The (canonically
normalized) inflaton,
\beq\dphi=\vev{J}\dph\>\>\>\mbox{with}\>\>\> \dph=\phi-M
\>\>\>\mbox{and}\>\>\>\vev{J}\simeq\sqrt{2Np}\label{dphi} \eeq
acquires mass, at the SUSY vacuum in \Eref{vevs}, which is given
by
\beq \label{msn} \msn=\left\langle\Ve_{\rm
pHI,\se\se}\right\rangle^{1/2}= \left\langle \Ve_{\rm
pHI,\sg\sg}/J^2\right\rangle^{1/2}\simeq{\ld M}/{\sqrt{Np}}\eeq
and lies in the intermediate energy scale. E.g., for $p=2$ we find
\beq 0.14\lesssim \msn/\ZeV \lesssim3.8 ~~\mbox{with}~~ 8\gtrsim
N\gtrsim 0.1.\label{resp2}\eeq
Obviously $\msn$ increases as $N$ decreases. Recall that
$1~\ZeV=10^{12}~\GeV$.

During the phase of its oscillations at the SUSY vacuum, $\dphi$
decays pertubatively  reheating the Universe at a reheat
temperature given by \cite{phi}
\beq\Trh= \left({40\over\pi^2g_{\rm
rh*}}\right)^{1/4}\lf\Gsn\mP\rg^{1/2}\>\>\>\mbox{with}\>\>\>\Gsn=\GNsn+\Ghsn\,.\label{Trh}\eeq
Also $g_{*}=228.75$ counts the MSSM effective number of
relativistic degrees of freedom. To compute $\Trh$ we take into
account the following decay widths:
\beqs\bea \label{Gn}
\GNsn&=&\frac{g_{iN^c}^2}{16\pi}\msn\lf1-\frac{4\mrh[
i]^2}{\msn^2}\rg^{3/2}\>\>\mbox{with}\>\>
g_{iN^c}=\sqrt{2}\ld_{iN^c}/{\vev{J}},\\
\label{Gh} \Ghsn&=&\frac{2}{8\pi}g_{H}^2\msn\>\>\>\mbox{with}\>\>
g_{H}={\lm}/{\sqrt{2}} \eea \eeqs which arise from the lagrangian
terms
\beqs \bea {\cal L}_{\dphi\to
\sni\sni}&=&-\frac12e^{K/2\mP^2}W_{{\rm B},N_i^cN^c_i}\sni\sni\ +\
{\rm h.c.}=g_{iN^c} \dphi\ \lf\sni\sni\ +\ {\rm h.c.}\rg
+\cdots,\\ {\cal
L}_{\dphi\to\hu\hd}&=&-e^{K/\mP^2}K^{SS^*}\left|W_{{\rm
B},S}\right|^2 =-g_{H} \msn\dphi \lf H_u^*H_d^*\ +\ {\rm
h.c.}\rg+\cdots\eea\eeqs
describing $\dphi$ decay into a pair of $N^c_{j}$ and $\hu$ and
$\hd$ respectively. Note that $N^c_{j}$ acquire masses
$\mrh[j]=\sqrt{2}\ld_{jN^c}M$ and we work in the so-called
\emph{$\sni$-basis}, where $\mrh[j]$ is diagonal, real and
positive.


For $\Trh<\mrh[i]$, the out-of-equilibrium decay of $N^c_{i}$
generates a lepton-number asymmetry (per $N^c_{i}$ decay),
$\ve_i$. The resulting lepton-number asymmetry is partially
converted through sphaleron effects into a yield of the observed
BAU \cite{phi}
\beq Y_B=-0.35\cdot{3\over2}\ {\Trh\over\msn}\mbox{$\sum_i$}
{\GNsn\over\Gsn}\ve_i,\label{Yb}\eeq
where $\ve_i$ are calculated as analyzed in \cref{unv}. The
expression above has to match with the observational value
\cite{act}
\beq \Yb=\lf8.75\pm0.085\rg\cdot10^{-11}.\label{bdata}\eeq
The validity of \Eref{Yb} requires that the $\dphi$ decay into a
pair of $\sni$'s is kinematically allowed for at least one species
of the $\sni$'s and also that there is no erasure of the produced
$Y_L$ due to $N^c_1$ mediated inverse decays and $\Delta L=1$
scatterings. These prerequisites are ensured if we impose
\beq\label{kin} 10\Trh\lesssim\mrh[1]\lesssim\msn/2.\eeq
The quantity $\ve_i$ can be expressed in terms of the Dirac masses
of $\nu_i$, $\mD[i]$, arising from the third term of $W$ in
\Eref{Whi}. Employing the seesaw formula we can then obtain the
light-neutrino mass matrix $m_\nu$ in terms of $\mD[i]$ and
$\mrh[i]$. As a consequence, nTL can be nicely linked to low
energy neutrino data \cite{valle}.

The required $\Trh$ in \Eref{Yb} must be compatible with
constraints on the gravitino ($\Gr$) abundance, $\Yg$, at the
onset of \emph{nucleosynthesis} ({\sf\small BBN}), which is
estimated to be \cite{kohri}
\beq\label{Ygr} \Yg\simeq 1.9\cdot10^{-13}~\Trh/\EeV ,\eeq
where we take into account only thermal production of $\Gr$, and
assume that $\Gr$ is much heavier than the MSSM gauginos.
Avoidance any disturbance to BBN entails \cite{kohri}
\beq  \label{grdata} \Yg\lesssim\left\{\bem
%
10^{-14}\hfill \cr
10^{-13}\hfill \cr \eem
\right.\>\>\>\mbox{for}\>\>\>\mgr\simeq\left\{\bem
0.69~\TeV\hfill \cr
10.6~\TeV\hfill \cr\eem
\right.\>\>\>\mbox{implying}\>\>\>\Trh\lesssim5.3\cdot\left\{\bem
%
0.01~\EeV\,,\hfill \cr
0.1~\EeV\,,\hfill \cr \eem
\right.\eeq
for the typical case where $\Gr$ decays with a tiny hadronic
branching ratio. The bounds above can be somehow relaxed in the
case of a short or very short-lived $\Gr$ -- cf., e.g.,
\cref{actIHI}.

\subsection{Results}\label{num}

Confronting with observations $Y_B$ and $\Yg$ -- see
\eqs{bdata}{grdata} -- which depend on $\msn$, $\Trh$, $\mrh[i]$
and $\mD[i]$'s  -- see \eqs{Yb}{Ygr} -- we can further constrain
the parameter space of the our model. We follow the bottom-up
approach detailed in \cref{unv}, according to which we find the
$\mrh[i]$'s by using as inputs the $\mD[i]$'s, $\mn[1]$ for
\emph{normal ordered} ({\sf\small NO}) \emph{neutrino masses}
$\mn[i]$'s, the two Majorana phases $\varphi_1$ and $\varphi_2$ of
the PMNS matrix, and the best-fit values for the remaining
neutrino oscillation parameters. Namely, we take as inputs the
recently updated best-fit values \cite{valle} -- cf. \cref{unv} --
on the neutrino oscillation parameters including IceCube IC24 with
Super Kamiokande (SK) atmospheric data. We consider only the
scheme of NO $\mn[i]$'s, which can become consistent with the
upper bound on the sum of $\mn[i]$'s
\beq\mbox{$\sum_i$} \mn[i]\leq0.082~{\eV},\label{sumnu}\eeq
induced from {\sf\small P-ACT-LB} data \cite{act} at 95\% c.l.
Namely, we take \cite{valle} $\Delta
m^2_{21}=7.49\cdot10^{-5}~{\rm eV}^2$ and $\Delta
m^2_{31}=2.513\cdot10^{-3}~{\rm eV}^2$ for the mass-squared
differences, $\sin^2\theta_{12}=0.308$,
$\sin^2\theta_{13}=0.02215$ and $\sin^2\theta_{23}=0.47$ for the
mixing angles, and $\delta=1.1788\pi$ for the CP-violating Dirac
phase.


\begin{figure}[!t]\vspace*{-.12in}
\hspace*{-.1in}
\begin{minipage}{8in}
\epsfig{file=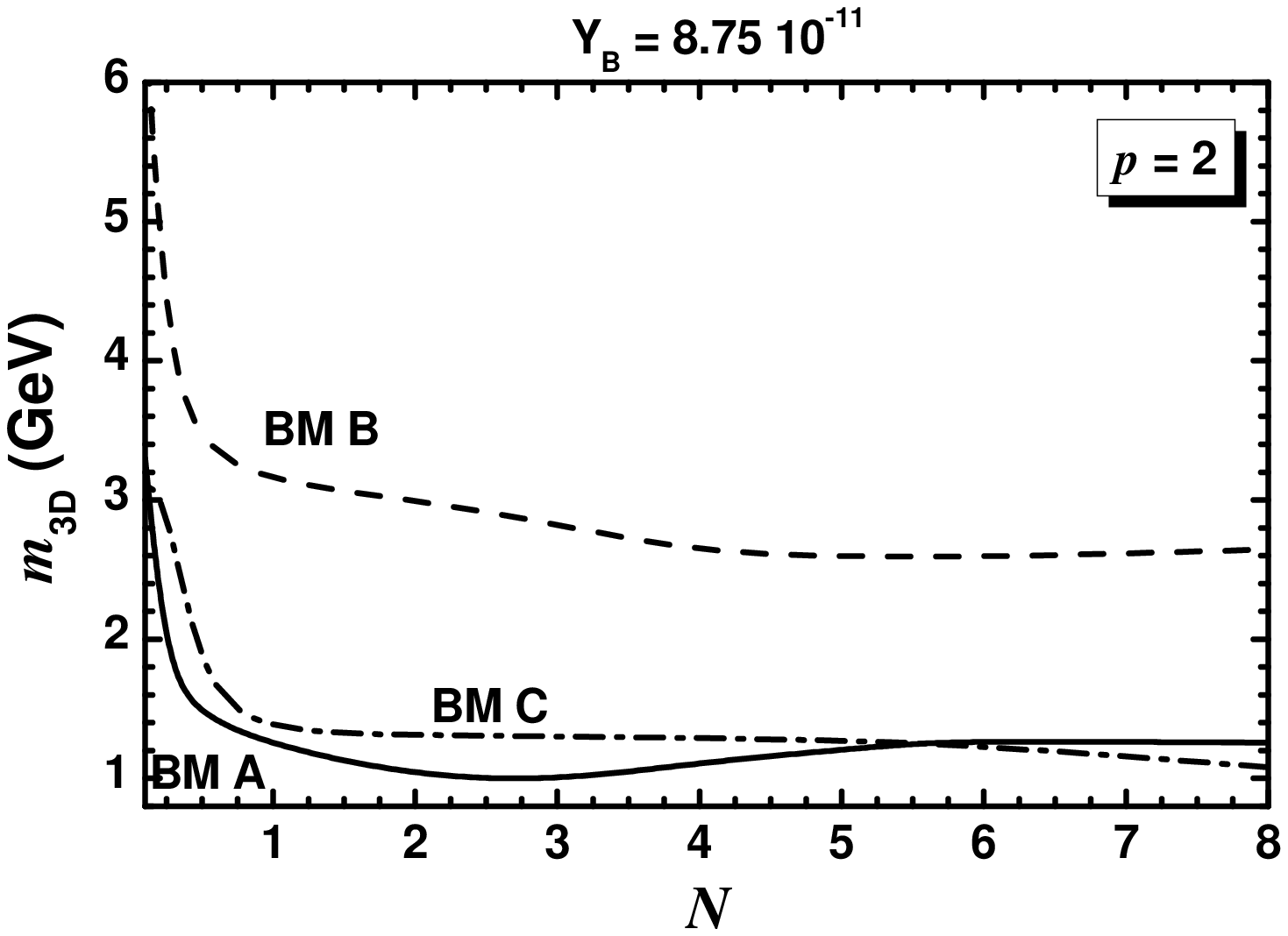,height=3.4in,angle=-90} \hspace*{-1.cm}
\epsfig{file=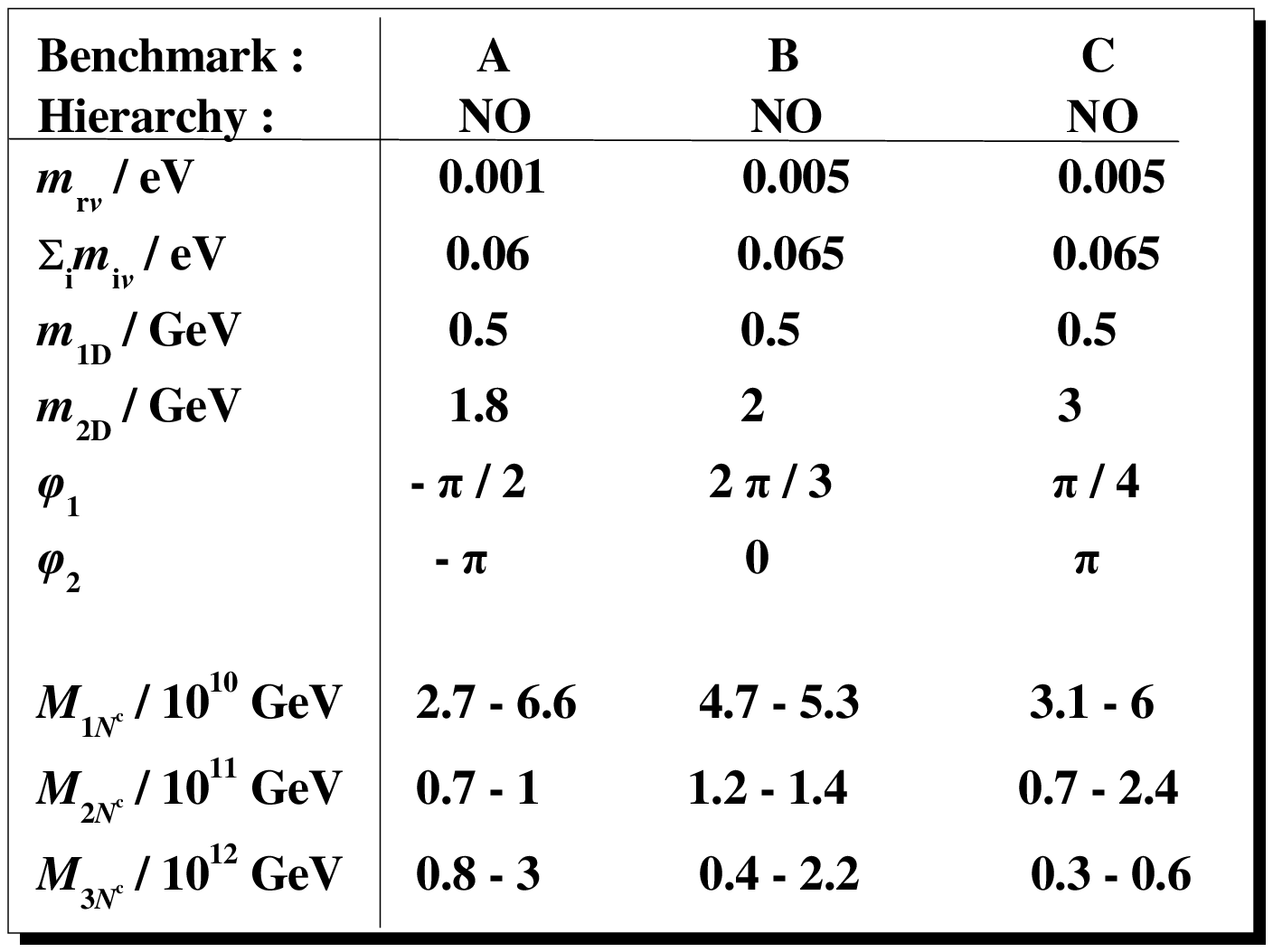,height=3.4in,angle=-90} \hfill
\end{minipage}
\hfill \caption{\sl\small  Contours in the $N-m_{\rm 3D}$ plane
yielding the central $Y_B$ in Eq.~(4.14) consistently with the
remaining inflationary and post-inflationary requirements for
$\mgr=10\mu=50~\TeV$, $\nst=1$ and the values of $m_{i\nu}$,
$m_{\rm 1D}$, $m_{\rm 2D}$, $\varphi_1$ and $\varphi_2$ which
correspond to benchmarks  A (solid line), B (dashed line) and C
(dot-dashed line) shown in the Table.}\label{fmD}
\end{figure}

Some representative results of our analysis are presented in
\fref{fmD} where we depict curves in the $N-\mD[3]$ plane which
yield the central $\Yb$ in \Eref{bdata} for $p=2$,
$\mu=0.1\mgr=5~\TeV$ and $\nst=1$ consistently with \Eref{Ntot} --
(\ref{Nmin}) and \Eref{grdata} -- note that the allowed $N$ values
are given in \Eref{resp2}. We use solid, dashed and dot-dashed
line when the remaining inputs -- i.e. $\mn[1]$, $\mD[1]$,
$\mD[2]$, $\varphi_1$, and $\varphi_2$ -- correspond,
respectively, to benchmarks A, B and C of the Table in \fref{fmD}.
Since the range of $Y_B$ in \Eref{bdata} is very narrow, the
$95\%$ c.l. width of these contours is negligible. The gauge
symmetry considered in work does not predict any particular Yukawa
unification pattern and so, the $\mD[i]$'s are free parameters.
This fact facilitates the fulfilment of the bounds in \Eref{kin}
since $\mD[1]$ affects heavily $\mrh[1]$. Care is also taken so
that the perturbativity of $\ld_{iN^c}$ holds, i.e.,
$\ld_{iN^c}^2/4\pi\leq1$. In all cases the inflaton $\dphi$ decays
into $N_i^c$'s  with $i=1$ and 2 with $\GNsn>\Ghsn$ and so the
ratios $\GNsn/\Gsn$ do not introduce a considerable reduction in
the derivation of $\Yb$. As we observe from the data in
\fref{fmD}, successful nTL requires $\mrh[i]$ and $\mD[3]$ in the
ranges $(0.01-1)~\ZeV$ and $(0.5-6)~\GeV$ respectively. As regards
the other quantities, in all we obtain
\beq
2.7\cdot10^{-2}\lesssim\Trh/{\EeV}\lesssim3\>\>\>\mbox{with}\>\>\>0.3\lesssim\lm/10^{-6}\lesssim9.4,\label{res3}\eeq
where the variation of $\lm$ is explained due to the variation of
$N$ -- see \Eref{mu}. Although we did not study the impact of a
variation of $\mu/\mgr$ to our results, we intuitively expect that
increasing $\mu/\mgr$ beyond its present value $0.1$ makes the
achievement of the correct $\Yb$ and $\Yg$ difficult. This is
because an increase of $\mu/\mgr$ implies an elevation of $\lm$ --
see \Eref{mu} -- and of $\Ghsn$ in \Eref{Gh}. The last effect
causes a reduction of the ratio $\GNsn/\Gsn$ which entails an
increase to $\Trh$ in order to readjust $\Yb$. However, the
increase on $\Trh$ increases $\Yg$ and renders the $\Gr$ problem
more acute. Therefore, our scheme primarily prefers
$\mu/\mgr\lesssim1$ -- cf. \cref{phi} -- which is defended by
various analysis \cite{klp, gabit, gmssm} of MSSM.


As a bottom line, nTL not only is a realistic possibility within
our setting but also it can be comfortably reconciled with the
$\Gr$ constraint even for $\mgr\sim1~\TeV$ as deduced from
\eqs{res3}{grdata}.

\section{Conclusions}\label{con}

We investigated the realization of pHI and nTL in the framework of
a $B-L$ extension of MSSM endowed with the condition that the GUT
scale is determined by the running of the three gauge coupling
constants of MSSM. Our setup is tied to the super-{} and \Kap s
given in Eqs.~(\ref{Whi}) and (\ref{tkis}). Prominent in this
setting is the role of the inflationary part of $K$, $\tks$, in
\Eref{tki} which is rational and exhibits a shift symmetry which
making it to vanish during pHI. Its free parameters $(p,N)$ can be
constrained by the observations -- see e.g. \Eref{res1}. Despite
$\tks$ does not enjoy a specific symmetry as in the case of
T-model HI \cite{sor,tmhi}, it assists in obtaining an excellent
match with \actc\ data -- see \fref{fig2}. In addition, our model
exhibits the following features: {\sf\small (i)} It assures
possibly detectable primordial gravitational waves for natural (of
order unity) $(p,N)$ values; {\sf\small (ii)} It inflates away
cosmological defects; {\sf\small (iii)} It offers a nice solution
to the $\mu$ problem of MSSM, provided that $\lm$ in \Eref{Whi} is
somehow small; {\sf\small (iv)} It allows for baryogenesis via nTL
compatible with $\Gr$ constraint and neutrino data. In particular,
we may have $\mgr\sim1~\TeV$, with the inflaton decaying mainly to
$N^c_1$ and $N^c_2$ -- we obtain $\mrh[i]$ in the range
$(10^{10}-10^{12})~\GeV$. It remains to introduce a consistent
soft SUSY-breaking sector to gain a more
precise completion.



\def\ijmp#1#2#3{{\emph{Int. Jour. Mod. Phys.}}
{\bf #1},~#3~(#2)}
\def\plb#1#2#3{{\emph{Phys. Lett.  B }}{\bf #1},~#3~(#2)}
\def\zpc#1#2#3{{Z. Phys. C }{\bf #1},~#3~(#2)}
\def\prl#1#2#3{{\emph{Phys. Rev. Lett.} }
{\bf #1},~#3~(#2)}
\def\rmp#1#2#3{{Rev. Mod. Phys.}
{\bf #1},~#3~(#2)}
\def\prep#1#2#3{\emph{Phys. Rep. }{\bf #1},~#3~(#2)}
\def\prd#1#2#3{{\emph{Phys. Rev.  D }}{\bf #1},~#3~(#2)}
\def\npb#1#2#3{{\emph{Nucl. Phys.} }{\bf B#1},~#3~(#2)}
\def\npps#1#2#3{{Nucl. Phys. B (Proc. Sup.)}
{\bf #1},~#3~(#2)}
\def\mpl#1#2#3{{Mod. Phys. Lett.}
{\bf #1},~#3~(#2)}
\def\arnps#1#2#3{{Annu. Rev. Nucl. Part. Sci.}
{\bf #1},~#3~(#2)}
\def\sjnp#1#2#3{{Sov. J. Nucl. Phys.}
{\bf #1},~#3~(#2)}
\def\jetp#1#2#3{{JETP Lett. }{\bf #1},~#3~(#2)}
\def\app#1#2#3{{Acta Phys. Polon.}
{\bf #1},~#3~(#2)}
\def\rnc#1#2#3{{Riv. Nuovo Cim.}
{\bf #1},~#3~(#2)}
\def\ap#1#2#3{{Ann. Phys. }{\bf #1},~#3~(#2)}
\def\ptp#1#2#3{{Prog. Theor. Phys.}
{\bf #1},~#3~(#2)}
\def\apjl#1#2#3{{Astrophys. J. Lett.}
{\bf #1},~#3~(#2)}
\def\n#1#2#3{{Nature }{\bf #1},~#3~(#2)}
\def\apj#1#2#3{{Astrophys. J.}
{\bf #1},~#3~(#2)}
\def\anj#1#2#3{{Astron. J. }{\bf #1},~#3~(#2)}
\def\mnras#1#2#3{{MNRAS }{\bf #1},~#3~(#2)}
\def\grg#1#2#3{{Gen. Rel. Grav.}
{\bf #1},~#3~(#2)}
\def\s#1#2#3{{Science }{\bf #1},~#3~(#2)}
\def\baas#1#2#3{{Bull. Am. Astron. Soc.}
{\bf #1},~#3~(#2)}
\def\ibid#1#2#3{{\it ibid. }{\bf #1},~#3~(#2)}
\def\cpc#1#2#3{{Comput. Phys. Commun.}
{\bf #1},~#3~(#2)}
\def\astp#1#2#3{{Astropart. Phys.}
{\bf #1},~#3~(#2)}
\def\epjc#1#2#3{{Eur. Phys. J. C}
{\bf #1},~#3~(#2)}
\def\nima#1#2#3{{Nucl. Instrum. Meth. A}
{\bf #1},~#3~(#2)}
\def\jhep#1#2#3{{\emph{JHEP} }
{\bf #1},~#3~(#2)}
\def\jcap#1#2#3{{\sl J. Cosmol. Astropart.
Phys. } {\bf #1},~#3~(#2)}
\def\jcapn#1#2#3#4{{\sl JCAP }{\bf #1}, no. #4, #3 (#2)}
\def\prdn#1#2#3#4{{\sl Phys. Rev. D }{\bf #1}, no. #4, #3 (#2)}
\newcommand{\arxiv}[1]{{\small\tt  arXiv:#1}}
\newcommand{\hepph}[1]{{\small\tt  hep-ph/#1}}
\def\prdn#1#2#3#4{{\sl Phys. Rev. D }{\bf #1}, no. #4, #3 (#2)}
\def\jcapn#1#2#3#4{{\sl J. Cosmol. Astropart.
Phys. }{\bf #1}, no. #4, #3 (#2)}
\def\epjcn#1#2#3#4{{\sl Eur. Phys. J. C }{\bf #1}, no. #4, #3 (#2)}

\end{document}